\documentstyle[12pt]{article}
\def\be{\begin{equation}}
\def\ee{\end{equation}}
\def\ba{\begin{eqnarray}}
\def\ea{\end{eqnarray}}
\def\ga{\mathrel{\raise.3ex\hbox{$>$\kern-.75em\lower1ex\hbox{$\sim$}}}}
\def\la{\mathrel{\raise.3ex\hbox{$<$\kern-.75em\lower1ex\hbox{$\sim$}}}}
\newcommand{\sect}[1]{\section{#1}\setcounter{equation}{0}}

\newcommand{\bi}[1]{\bibitem{#1}}

\begin{document}
\baselineskip=16pt
\begin{titlepage} 
\rightline{CERN--TH/2001-109}
\rightline{UMN--TH--2001/01}
\rightline{TPI--MINN--01/15}
\rightline{IHES/P/01/16}
\rightline{hep-ph/0104177}
\rightline{April 2001}  
\begin{center}

\vspace{0.5cm}

\large {\bf A 6-D Brane World Model
}
\vspace*{5mm}
\normalsize

{\bf  Panagiota Kanti$^1$, Richard Madden$^2$} and {\bf Keith A. Olive$^3$}

\smallskip 
\medskip 
$^1${\it Scuola Normale Superiore, Piazza dei Cavalieri 7,\\
I-56126 Pisa, Italy}

$^2${\it Institut des Hautes \'Etudes Scientifiques, 91440
Bures-sur-Yvette, France} 
 
$^3${\it Theoretical Physics Institute, School of Physics and
Astronomy,\\  University of Minnesota, Minneapolis, MN 55455, USA} 

\smallskip 
\end{center} 
\vskip0.6in 
 
\centerline{\large\bf Abstract}

We consider a 6D space-time which is periodic in one of the extra dimensions and
compact in the other. The periodic direction is defined by two 4-brane boundaries.
Both static and non-static exact solutions, in which the internal
spacetime has constant radius of curvature, are derived. In the case of
static solutions, the brane tensions must be tuned as in the 5D
Randall-Sundrum model, however, no additional fine-tuning is necessary
between the brane tensions and the bulk cosmological constant. By further
relaxing the sole fine-tuning of the model, we derive non-static solutions,
describing de Sitter or Anti de Sitter 4D spacetimes, that allow for the
fixing of the inter-brane distance and the accommodation of pairs of
positive-negative and positive-positive tension branes.  Finally, we
consider the stability of the radion field in these configurations by
employing small, time-dependent perturbations around the background
solutions. In analogy with results drawn in 5 dimensions, the solutions
describing a de Sitter 4D spacetime turn out to be unstable while those
describing an Anti de Sitter geometry are shown to be stable.

\vfill
\vskip 0.15in
\leftline{CERN--TH/2001-109}
\leftline{April 2001}
\end{titlepage}
\baselineskip=18pt

\sect{Introduction}

It would be an understatement to say that the possibility of resolving
the hierarchy problem in models with a warped extra dimension \cite{RS1} has received
considerable attention over the last two years.  Indeed, brane world models
have dominated the literature in high energy theory.
The reason lies in its simplicity. By postulating the existence of 
two 3-branes with non-zero tensions, separated along the extra dimension by a distance
$L$, in the background of a non-zero (negative) cosmological constant, one
finds a simple solution for the scale factor along the extra dimension,
$a(y)$,  which is exponential.  Thus, length scales (and hence mass scales)
on one brane  are exponentially enhanced (or suppressed) relative to the
other. A mass hierarchy naturally arises between the two branes which can be
labelled the Planck and weak branes respectively.

Of course, there is a price to pay for this simplicity.
First, as it is well known, the tensions of the two 3-branes, must be fine-tuned
so that $\Lambda_1 = -\Lambda_2$. Second, these tensions must be tuned to 
the bulk cosmological constant, $\Lambda_B$, in order to produce a static solution.
The origins of these  fine-tunings come about when one considers
static solutions to the 5D equations of motion.  The scale factor in the
extra direction takes the form $a(y) = e^{-k y}$ and the equations of motion
require $k^2 = -\kappa^2_5 \Lambda_B/6$, where $\kappa^2_5$ is the 5D Newton's
constant. We are, therefore, led
to an Anti de Sitter 5D spacetime with $\Lambda_B < 0$.
By putting branes in the theory, and requiring that the warp factor 
exhibits periodic behavior along the extra dimension, we obtain the so called 
{\it jump} conditions which give $[a'(y)]_i/a_i = -\kappa^2_5 \Lambda_i/3$,
where $[a']$ represents the difference in $a'$ on the two sides of the brane. 
For one brane say with positive tension placed at the
origin ($y=0$), we see that $k = \kappa^2_5 \Lambda_1/6$, and for the second brane,
located at an
undetermined distance $L$, one finds $k = -\kappa^2_5 \Lambda_2/6$.  Thus, we arrive
at the conditions $\Lambda_1 = -\Lambda_2 = \sqrt{6 \Lambda_B/\kappa^2_5}$. 
Finally, the distance L is
chosen so as to resolve the hierarchy problem by noting that masses scale as $a(y)$.
For other recent attempts at solving the hierarchy problem with extra dimensions
see \cite{5D,kop2}. 

Several extra-dimensional attempts at resolving the hierarchy, or the
cosmological constant, problem have considered six- or higher-dimensional
models \cite{6D1, 6D2}. Space-times with more than 1 extra dimension can
allow for solutions with most appealing features, particularly in spacetimes
where the curvature of the internal space is non-zero. These solutions,
exhibiting either spherical or cylindrical symmetry with respect to the extra
coordinates, can accommodate an exponential dependence on one of the extra
coordinates, thus, resembling the 5D RS mechanism for the resolution of the
hierarchy problem. In addition, it turns out that such space-times can play
an important role in relaxing the degree of fine-tuning in the RS models
\cite{6D2}. Finally, these models can provide a framework in the context
of which the stabilization of the radion field naturally takes place:
for example, in Ref. \cite{zhuk}, it was shown that in space-times with a
constant  spatial curvature of the internal dimensions, one can find
solutions with a global minimum in the effective theory for the radion
field. 

In this paper, we look for solutions to the 6D equations of motion based
on an internal space of constant curvature. We first present an exact
static solution where the warp factor depends on both extra coordinates
and, hence, does not exhibit any spherical or cylindrical symmetry. 
The dependence on one of the two extra coordinates is a purely exponential
one thus resembling the profile of the warp factor in the case of the 5D
RS model. In analogy with RS1, the space-time contains two 4-brane
boundaries with equal and opposite brane tensions.  This configuration
requires the same fine-tuning that exists in RS1 model due to the {\it jump}
conditions imposed at the boundaries. However, the solution does not contain
the additional fine-tuning between the brane tensions and the bulk
cosmological constant which is replaced by the fixing of the size of the
extra dimension along which the 4-branes extend. The inter-brane distance
along the remaining extra dimension remains arbitrary and it may be fixed
only through the introduction of an additional mechanism for the
stabilization of the radion field \cite{kkop2,randall3}.

We then proceed to derive non-static solutions in the context of the same
model. In this case, the exponential behaviour along one of the two extra
coordinates changes to $cosh$ or $sinh$-like allowing for the
accommodation of pairs of branes with positive tensions or positive-negative, 
respectively. The {\it jump} conditions lead to the fixing of the locations
of the two branes along the same dimension and the fine-tuning between the
brane tensions disappears rendering this solution totally free of any
fine-tuning.

Both of the above solutions, static or non-static, have been derived under
the assumption that the extra space-time remains static. We formulate
an ``extremization" constraint that may serve as a consistency check
for any 6D solution with a constant or non-constant radion field. We
finally perform a stability analysis around the solutions with a
constant ``extra'' scale factor in order to check their stability
under small time-dependent perturbations. We find, in agreement
with similar results derived for 5D spacetimes \cite{Gen} - \cite{CF}, that
the system of two Minkowski 4D subspaces has a vanishing radion mass, a pair
of two de Sitter ones has a negative mass squared while the system of two
Anti de Sitter 4D subspaces has a positive mass squared.

In the next section, we present the model and derive the exact static
solution in section 3. We show explicitly how the correlation between
the brane tensions and the bulk cosmological constant is replaced by
the fixing of the size of one of the two extra dimensions.  In section 4,
we show how the relaxation of the fine-tuning between the brane tensions
leads to de Sitter or Anti de Sitter expansions in the 4D space-time
in analogy with 5D models. The ``extremization'' constraint and the
stability analysis of our solutions are discussed in section 5. 
Finally, we present our conclusions in section 6.

\sect{The general framework}

Let us start by presenting the theoretical framework and geometrical
set-up of our model. We first write down the action, that 
describes the gravitational theory of a 6-dimensional spacetime filled
with a bulk cosmological constant, as\footnote{Throughout this paper, we
follow Wald's conventions \cite{wald}: The metric signature is
$\eta_{MN}=(-,+,...,+)$ and the Riemann tensor is defined as
$R^\sigma_{\,\,\rho\mu\nu}=\partial_\mu \Gamma^\sigma_{\rho\nu}-
\partial_\nu\Gamma^\sigma_{\rho\mu} + ...\ $.}
\be
S=-\int \,d^6x\,\sqrt{-G_6}\,\biggl\{-\frac{R_{(6)}}{2\kappa_6^2}
+ \Lambda_B\biggr\}\,,
\label{action}
\ee
where $\kappa_6^2=8\pi/M_6^4$. The line-element of the 6-dimensional
spacetime is assumed to be of the form
\be
ds^2=a^2(t,\theta, \varphi)\,\eta_{\mu\nu}\,dx^\mu dx^\nu + b^2(t)
\,[d\theta^2 + f^2(\theta) \,d\varphi^2]\,.
\label{metric}
\ee
In the above, $x^\mu$ and ($\theta, \phi$) denote the coordinates along
the usual four and two extra, all (initially) non-compact, dimensions,
respectively. The function $a(t,\theta, \varphi)$ represents the warp
factor multiplying the 4-dimensional line-element, $b(t)$ denotes the scale
factor that determines the dynamics of the 2-dimensional extra spacetime
while $f(\theta)$ parameterizes its internal curvature. For reasons that 
will become clear shortly, we will often refer to the
$\theta$-dimension as the ``longitudinal'' one and to the $\varphi$-dimension
as the ``transverse" one. 

In this paper, we will focus on the determination of static and non-static
solutions under the assumption that the scale factor along the extra dimensions
remains always constant. As noted in the Introduction, we are not assuming
any specific mechanism for radion stabilization which would generate additional 
stress-energy terms \cite{kkop2,randall3}. We will instead assume that the
internal curvature of the 2-dimensional extra spacetime will naturally lead
the system to solutions with a constant radion field that correspond to a
minimum of the radion effective potential as in Ref. \cite{zhuk}.
The outcome of this attempt is not however straightforward: in \cite{zhuk},
it was assumed that $a$ and $b$ are functions of the coordinates $x^\mu$ only.
None of the scale factors depend on the internal coordinates $\theta$ and
$\phi$ and, thus, their results are not directly applicable to any warped
brane model. 

Upon variation with respect to the 6D metric tensor, the above action
and metric ansatz leads to the following Einstein's equations in the bulk
\ba
&~&  \hspace*{-1cm} G_{00} = \frac{\dot b^2}{b^2} 
+\frac{6 \dot b \dot a}{ba} +\frac{3\dot a^2}{a^2}-
\frac{a^2}{b^2}\,\frac{f''}{f} -
\frac{3}{b^2 f^2}\,\Bigl(\frac{\partial a}{\partial \varphi}\Bigr)^2
-\frac{3a}{b^2 f^2}\,\frac{\partial^2 a}{\partial \varphi^2}
\nonumber \\[2mm] &~&  \hspace*{5.5cm} 
-\frac{3 a f'}{b^2 f}\,\frac{\partial a}{\partial \theta}
- \frac{3}{b^2}\,\Bigl(\frac{\partial a}{\partial \theta}\Bigr)^2
-\frac{3a}{b^2}\,\frac{\partial^2 a}{\partial \theta^2}=
\kappa^2_6\,a^2 \Lambda_B\,,\label{geq00} \\[4mm]
&~& \hspace*{-1cm}  G_{ii} = -\frac{\dot b^2}{b^2} -
\frac{2 \ddot b}{b} 
-\frac{2 \dot b \dot a}{ba} +\frac{\dot a^2}{a^2}
-\frac{2 \ddot a}{a} + \frac{a^2}{b^2}\,\frac{f''}{f} +
\frac{3}{b^2 f^2}\,\Bigl(\frac{\partial a}{\partial \varphi}\Bigr)^2
+\frac{3a}{b^2 f^2}\,\frac{\partial^2 a}{\partial \varphi^2}
\nonumber \\[2mm] &~& \hspace*{5.2cm}
+\frac{3 a f'}{b^2 f}\,\frac{\partial a}{\partial \theta}
+\frac{3}{b^2}\,\Bigl(\frac{\partial a}{\partial \theta}\Bigr)^2
+\frac{3a}{b^2}\,\frac{\partial^2 a}{\partial \theta^2}=
-\kappa^2_6\,a^2 \Lambda_B\,,\label{geqii} \\[4mm]
&~& \hspace*{-1cm} G_{55} = -\frac{b \ddot b}{a^2}
-\frac{2 b \dot b \dot a}{a^3} -\frac{3b^2 \ddot a}{a^3}
+ \frac{6}{a^2 f^2}\,\Bigl(\frac{\partial a}{\partial \varphi}\Bigr)^2
+\frac{4}{a f^2}\,\frac{\partial^2 a}{\partial \varphi^2}
\nonumber \\[2mm] &~& \hspace*{7.2cm}
+\frac{4 f'}{a f}\,\frac{\partial a}{\partial \theta}+
\frac{6}{a^2}\, \Bigl(\frac{\partial a}{\partial \theta}\Bigr)^2=
-\kappa^2_6\,b^2 \Lambda_B\,, \label{geq55}\\[4mm]
&~& \hspace*{-1cm} G_{66} =-\frac{b f^2 \ddot b}{a^2}
-\frac{2 b f^2 \dot b \dot a}{a^3} -\frac{3b^2 f^2 \ddot a}{a^3}
+ \frac{6}{a^2}\,\Bigl(\frac{\partial a}{\partial \varphi}\Bigr)^2
+ \frac{6 f^2}{a^2}\, \Bigl(\frac{\partial a}{\partial \theta}\Bigr)^2
\nonumber \\[2mm] &~& \hspace*{8.6cm}  
+ \frac{4 f^2}{a}\, \frac{\partial^2 a}{\partial \theta^2}=
-\kappa^2_6\,b^2 f^2 \Lambda_B \,, \label{geq66}\\[4mm]
&~& \hspace*{-1cm} G_{05} = 
\frac{4 \dot b}{a b}\,\frac{\partial a}{\partial \theta} +
\frac{3 \dot a}{a^2}\,\frac{\partial a}{\partial \theta}
-\frac{3}{a}\,\frac{\partial^2 a}{\partial t\,\partial \theta}=0\,,
\label{geq05}\\[4mm]
&~& \hspace*{-1cm} G_{06} =
\frac{4 \dot b}{a b}\,\frac{\partial a}{\partial \varphi} +
\frac{3 \dot a}{a^2}\,\frac{\partial a}{\partial \varphi}
-\frac{3}{a}\,\frac{\partial^2 a}{\partial t\,\partial \varphi}=0\,,
\label{geq06}\\[4mm]
&~& \hspace*{-1cm} G_{56} = 
\frac{4 f'}{af}\,\frac{\partial a}{\partial \varphi}
-\frac{4}{a}\,\frac{\partial^2 a}{\partial \theta \,\partial \varphi}=0\,.
\label{geq56}
\ea

As we mentioned above, both extra dimensions are initially non-compact. Here, we
choose to compactify along the $\varphi$-coordinate by
introducing two 4-branes at the points $\varphi=\varphi_1$ and 
$\varphi=\varphi_2$ (and impose periodic boundary conditions). The brane energy-momentum
tensors can be written as
\be
T^{M\,(i)}_N= \frac{\delta(\varphi-\varphi_i)}{bf^2}\,
{\rm diag}\Bigl(-\Lambda_i, -\Lambda_i, -\Lambda_i, -\Lambda_i,
-\tilde \Lambda_i, 0\Bigr)\,,
\label{tension}
\ee
where $i=1, 2$ denotes the two branes. Note that the requirement of satisfying
6D energy-momentum conservation, $T^{M\,(i)}_{N;M} = 0$, together with
the {\it jump} conditions
(discussed shortly) forces one to introduce 1) an ``inhomogeneity''
reflected in the prefactor of eq. (\ref{tension}) and 2) an ``anisotropy" 
such that $T^{5\,(i)}_5 \ne T^{k\,(i)}_k$ for $k=0,...,3$.
On the other hand, as
expected, the energy-momentum tensor in the bulk is dominated by
the presence of a smoothly distributed bulk cosmological constant, i.e.
\be
T^{M\,(B)}_N= {\rm diag}\Bigl(-\Lambda_B, -\Lambda_B, -\Lambda_B, -\Lambda_B,
-\Lambda_B, -\Lambda_B\Bigr)\,,
\label{bulk}
\ee
whose sign remains arbitrary at this point. Note that, in the absence of a 
radion potential, we do not include an additional contribution to
the (55) or (66)-component here.

Under the assumption that the two
4-branes are infinitely thin, the discontinuity in the first derivative
of the metric tensor along the $\varphi$-coordinate creates a
$\delta$-function contribution to its second derivative. The
{\it jump} conditions \cite{jump}, that follow by matching the coefficients 
of the $\delta$-functions on both sides of Einstein's equations,
provide constraints on the first derivatives of the warp factor
at the location of the branes. One may easily conclude, from the
absence of any delta-function source on the right-hand side of the
(66)-component of Einstein's equations, eq. (\ref{geq66}), that
there is no discontinuity in the first derivative of the warp factor
along the $\theta$-coordinate due to the smooth distribution of
energy along the longitudinal extra dimension. 
Then, from eqs. (\ref{geq00}) and (\ref{geq55}), which contain
second derivatives with respect to $\varphi$, we obtain respectively the
constraints
\be
\frac{[\partial_\varphi a]}{b\,a}\biggl|_{\varphi=\varphi_i}=
-\frac{\kappa^2_6}{3}\,\Lambda_i\,, \qquad \qquad
\frac{[\partial_\varphi a]}{b\,a}\biggl|_{\varphi=\varphi_i}=
-\frac{\kappa^2_6}{4}\,\tilde \Lambda_i\,.
\label{jump1}
\ee
{}From the above conditions, it is obvious that the (ii) and
(55)-components of the energy-momentum tensor of each brane must
satisfy the relation:
$\tilde \Lambda_i=4\Lambda_i/3$. Thus, the brane tension along the ``extra"
$\theta$-coordinate is clearly distinguished from the remaining four,
as pointed out below eq. (\ref{tension}). Similar ``anisotropic'' brane
tensions can be found elsewhere \cite{6D2} in higher dimensional models.

As noted above, the inhomogeneous prefactor in eq. (\ref{tension}) comes
from the energy-momentum conservation constraint on $T^{M\,(i)}_N$. Let us
assume for the moment that $T^{M\,(i)}_N$ is of the form
$\Bigl(g(\theta)\,\delta(\varphi-\varphi_i)/b\Bigr)\,{\rm diag}
(-\Lambda_i, -\Lambda_i, -\Lambda_i, -\Lambda_i,-\tilde \Lambda_i, 0\Bigr)$. 
We have allowed $g$ to be arbitrary and we
have retained the common normalization $(1/b)$. The $N=0$
conservation equation is identically zero for all solutions with $\dot b = 0$,
which includes all of the solutions we consider.
Note that this constraint on $b$ stems from the anisotropy $\Lambda_i \ne
\tilde \Lambda_i$. The $N = 1,2,3$ equations are trivial. The $N = 5$ equation 
reduces to 
\be
-(g' + 4{a'\over a} + {f'\over f})\,\tilde \Lambda_i + 4{a'\over a} \Lambda_i
=0\,,
\ee
with all derivatives taken with respect to $\theta$. 
With the anisotropy relation derived from the {\it jump} conditions this becomes
\be
(g' + {a'\over a} + {f'\over f})\,\tilde \Lambda_i = 0\,.
\ee
As we will show in the next section, our solution further requires
${a'\over a} = {f'\over f}$, from which it can be easily seen that
$g = 1/f^2$. We also note that without this tension prefactor the
{\it jump} conditions would contain an explicit factor of $f(\theta)$
necessitating a constant $f$ and hence a flat internal space. 

\sect{Static 6D solutions}

In this section, we focus on the derivation of an exact 6-dimensional static
solution with an exponential warp factor and a constant scale factor,
$b(t)=b_0$. In this case, the non-vanishing
components of Einstein's equations (after dropping all terms with time
derivatives) take the form
\ba
\frac{f''}{f} +
\frac{3}{a^2 f^2}\,\Bigl(\frac{\partial a}{\partial \varphi}\Bigr)^2
+\frac{3}{a f^2}\,\frac{\partial^2 a}{\partial \varphi^2}
+\frac{3 f'}{a f}\,\frac{\partial a}{\partial \theta}
+ \frac{3}{a^2}\,\Bigl(\frac{\partial a}{\partial \theta}\Bigr)^2
+ \frac{3}{a}\,\frac{\partial^2 a}{\partial \theta^2} &=&
-\kappa^2_6\,b_0^2 \Lambda_B\,, \label{eq1}\\[4mm]
\frac{6}{a^2 f^2}\,\Bigl(\frac{\partial a}{\partial \varphi}\Bigr)^2
+\frac{4}{a f^2}\,\frac{\partial^2 a}{\partial \varphi^2}
+\frac{4 f'}{a f}\,\frac{\partial a}{\partial \theta}+
\frac{6}{a^2}\, \Bigl(\frac{\partial a}{\partial \theta}\Bigr)^2 &=&
-\kappa^2_6\,b_0^2 \Lambda_B\,, \label{eq2} \\[4mm]
\frac{6}{a^2 f^2}\,\Bigl(\frac{\partial a}{\partial \varphi}\Bigr)^2
+ \frac{6}{a^2}\, \Bigl(\frac{\partial a}{\partial \theta}\Bigr)^2
+ \frac{4}{a}\, \frac{\partial^2 a}{\partial \theta^2}&=&
-\kappa^2_6\, b_0^2 \Lambda_B\,, \label{eq3} \\[4mm]
\frac{4 f'}{af}\,\frac{\partial a}{\partial \varphi}
-\frac{4}{a}\,\frac{\partial^2 a}{\partial \theta \, \partial \varphi}&=& 0
\,. \label{eq4}
\ea

We assume that the dependence of the warp factor on the two
extra coordinates can be written in a factorized form: 
$a (\theta, \varphi)= \Theta(\theta)\,\Phi (\varphi)$. 
Eq. (\ref{eq4}), then, provides a very strong constraint on the functions
$\Theta(\theta)$ and $f(\theta)$ leading to the relation
$\Theta(\theta)= A_\theta\,f(\theta)$, where $A_\theta$ is a constant.
The difference of eqs. (\ref{eq2}) and (\ref{eq3}), in turn, leads
to the result
\be
\frac{\Phi''}{\Phi}=f f''- f'^2=\omega^2\,,
\label{rear1}
\ee
where $\omega$ is again an arbitrary constant. The above equation allows
us to write the general solution for $\Phi(\phi)$ as a linear combination
of increasing and decreasing exponentials. However, eq. (\ref{eq3}), which
may be written as 
\be
6\Bigl(\frac{\Phi'}{\Phi}\Bigr)^2 + 6f'^2 + 4 f f'' =
-\kappa^2_2\, b_0^2 f^2 \Lambda_B\,,
\label{rear2}
\ee
restricts the form of the solution by demanding  $\Phi'/\Phi$ to
be a constant as well. As a result, we may write the solution for $\Phi$
in the form
\be
\Phi(\varphi) = A_\varphi\,e^{\pm \omega \varphi}\,,
\label{solphi}
\ee
where $A_\phi$ an integration constant. Eq. (\ref{solphi}), together with
eqs. (\ref{rear1})-(\ref{rear2}),  brings the equation for the remaining
unknown metric function, $f(\theta)$, to the form
\be
f'^2 - \lambda^2 f^2 + \omega^2 =0\,,
\label{eqf2}
\ee
whose general solution can be written as
\be
f(\theta)=\frac{1}{4\lambda}\,[e^{\pm \lambda (\theta-\theta_0)}
+ 4 \omega^2 e^{\mp \lambda (\theta-\theta_0)}]\,,
\label{general}
\ee
where
\be
\lambda^2 = - \frac{\kappa^2_6 b_0^2}{10}\,\Lambda_B\,.
\label{lambda}
\ee
Note that the parameter $\lambda$ can take real or imaginary values for
a negative or positive, respectively, bulk cosmological constant. We will
comment on these two different options shortly. 

We can furthermore easily check that the combination of eqs.
(\ref{rear1})-(\ref{rear2}) trivially satisfies the remaining equation
(\ref{eq1}). We may, therefore, write the full solution for the warp factor,
that multiplies the line-element of the 4-dimensional space-time, as
\be
a(\theta, \varphi)= \frac{a_0}{4\lambda}\,e^{\pm \omega \varphi}\,
[e^{\pm \lambda (\theta-\theta_0)}
+ 4 \omega^2 e^{\mp \lambda (\theta-\theta_0)}]\,,
\label{warp}
\ee
where $a_0$ stands for the product $A_\varphi\,A_\theta$. For simplicity,
we may choose to place one 4-brane at the point $\varphi=0$ and the second
one at $\varphi=L$. Then, by imposing the condition
$a(\theta=\theta_0, \varphi=0)=1$, the integration constant $a_0$ may be
fixed in terms of $\lambda$ and $\omega$.

Clearly, the $\varphi$-coordinate corresponds to a non-compact extra dimension
as the conformal factor increases, or decreases, exponentially. The 
similarity with the extra fifth dimension of the Randall-Sundrum model
is striking: the size of this dimension can become finite only by  
introducing two branes (3-branes in the case of the RS1 model, 4-branes in
this case) at two different points along the transverse $\varphi$-dimension. 
Then, the inter-brane distance defines the size of the extra dimension.
The monotonic behaviour of the scale factor in the $\phi$ direction, calls
for the introduction of a pair of branes with positive and negative tension,
as discussed earlier. We will show that this is indeed the case
at the end of this section. Single-brane configurations could be also
accommodated in our analysis by
sending the second brane to an infinite distance from the first one.

At this point, we can easily derive the relation between the internal 2D
curvature and the bulk cosmological constant. 
The 2-dimensional curvature scalar is 
\be
R_{(2)}= R^5_5 + R^6_6 = -\frac{2}{b_0^2}\,\frac{f''}{f}\,.
\label{2dcurv}
\ee
{}From the solution (\ref{general}), we further see that
$\frac{f''}{f} = \lambda^2$, so that
\be
R_{(2)}= \frac{\kappa^2_6}{5}\,\Lambda_B\,.
\label{2dlb}
\ee
The behaviour of the warp factor along the $\theta$-dimension, and
subsequently the topology of this dimension, is strictly defined by the
sign of the $\lambda^2$ parameter which determines the sign of the two-space
curvature through (\ref{2dcurv}), or equivalently through eq.
(\ref{lambda}), by the sign of the bulk cosmological constant. We now
distinguish the two cases:\\[3mm]
\underline{{\bf (A): $\lambda^2>0$.}} This case corresponds, through
eq. (\ref{lambda}), to a negative bulk cosmological constant and to a
negatively curved two-dimensional extra space-time
\be
\Lambda_B < 0\,, \qquad R_{(2)} < 0\,.
\ee
{}From the expression (\ref{general}), we see that the function
$f(\theta)$ is characterized by the existence of a minimum at
\be
\theta_{min}=\theta_0 \pm \frac{1}{\lambda}\,\ln 2\omega\,,
\label{size}
\ee
under the assumption that $f(\theta)$ is symmetric under the
transformation $\theta \leftrightarrow -\theta$. Identifying the two
minima (for simplicity, we may set $\theta_0=0$), we can compactify this
extra dimension. Then, the quantity $2\lambda\,\theta_{min}$ is the
physical size of the extra dimension, $2(b_0\,\theta_{min})$, in units of
$M_6^{-1}$. 

If we allow the two symmetric branches of the function $f(\theta)$ to
meet at $\theta=0$, a cusp is inevitably created. One might be tempted
to introduce an infinitely-thin 3-brane at this point creating a set-up that
resembles the 5D single-brane configurations presented in Ref.~\cite{kkop2}. 
However, the absence of any discontinuity in the first and second derivative
of the warp factor with respect to the $\theta$-coordinate does not allow
for the introduction of any infinitely-thin 3-brane in the model. The only
allowed configuration is the one where each one of the 4-branes defining
the size of the $\varphi$-dimension is a thick 3-brane extending along the
$\theta$-dimension similar to the one describing in Refs. \cite{kkop2,kkop1}.
In that case, the warp factor and its derivatives with respect to the
$\theta$-coordinate are everywhere well defined as demanded. Introducing
a thick 3-brane along the $\theta$-dimension will not spoil the solution
(\ref{warp}) found above: it would merely render it as the solution
outside the ``wall'' of the 3-brane where the minimum must take place. 

\underline{{\bf (B): $\lambda^2<0$.}} This case arises under the assumption
that the bulk cosmological constant takes a positive value. The
internal curvature of the ($\varphi, \theta$)-submanifold, in this case, is
also positive according to the definition (\ref{2dcurv}). Since
$\lambda^2<0$, we may write $\lambda = i \tilde \lambda$, where
$\tilde \lambda$ is a real number. It turns out that for special values of
the parameter $\omega$, the $\theta$-dependent part of the solution for the
warp factor (\ref{warp}) becomes periodic, and thus spontaneously compactified
without the need for the introduction of any thin or thick 3-branes. However,
in each case the resulting form of the line-element of the 6D spacetime
contradicts basic assumptions of this analysis.

Thus, if we choose $\omega^2 = 1/4$, the solution for $f(\theta)$ is given
in terms of a $cos$-type function. However, the presence of the coefficient
$1/(4\lambda)$ in eq. (\ref{general}) renders this metric function 
imaginary leading to:
\be
ds^2=a^2(\theta, \varphi)\,(-dt^2 + dx^2 + dy^2 + dz^2) +
b_0^2[d\theta^2 - \tilde f(\theta)^2 d\varphi^2]\,,
\ee
where we have set $f(\theta)=i \tilde f(\theta)$. As a result, the character
of the $\varphi$-dimension changes from space-like to time-like which is
in disagreement with our argument that the $\varphi$-dimension plays the
role of the extra transverse dimension of the RS model.
If, alternatively, we choose $\omega^2 = -1/4$, the function
$f(\theta)$ comes out to be proportional to a $sin$-type
function, however, in this case, $\omega$ itself comes out to
be imaginary. If we perform the following coordinate
transformation
\be
\varphi \rightarrow 4H \tilde t\,, \qquad \qquad
t \rightarrow \frac{i \tilde \varphi}{4H}\,,
\ee
the 6D line-element takes the form
\be
ds^2= \sin^2[\lambda\,(\theta-\theta_0)]\,\biggl\{ -d \tilde t^2
+ e^{2 H \tilde t}\Bigl(dx^2 + dy^2 + dz^2 + \frac{d\tilde\varphi^2}
{16 H^2}\Bigr)\biggr\} + b_0^2 d\theta^2
\ee
where $H=\tilde \lambda/(2b_0)$ and where, for simplicity, we have set
$a_0=2\tilde \lambda$ in eq. (\ref{warp}). The above line-element describes
a 6-dimensional spacetime whose 4 spatial dimensions are inflating.
This is clearly in contradiction with our main goal to derive solutions
with static extra dimensions.

We finally turn to the {\it jump} conditions that will help us determine
the $\omega$ parameter of the solution as well as the number of fine-tunings
that the model demands. 
Substituting the solution for $\Phi(\varphi)$ (\ref{solphi}) into the
{\it jump} conditions (\ref{jump1}), we obtain the result
\be
\pm \frac{3\omega}{\kappa^2_6\,b_0} = -\Lambda_1=\Lambda_2\,.
\label{jump3}
\ee
The same fine-tuning between the two brane tensions, that was necessary for
the RS solution to be consistent with the boundary conditions, appears
also in our model. Note, however, that the second RS fine-tuning, between
each brane tension and the bulk cosmological constant is absent. The
value of $\Lambda_B$ defines the parameter $\lambda$ while $\omega$
is defined in terms of the brane tensions. {\it No relation} between
 $\lambda$ and $\omega$ exists in our model, and thus, $\Lambda_B$
and $\Lambda_i$ remain uncorrelated. Nevertheless, since only one of the
two brane tensions has been fixed so far, another fundamental parameter
of our model, if not the second brane tension or the bulk cosmological
constant, should be fixed instead, in this case. Through eq. (\ref{size}),
the physical size of the $\theta$-dimension is given in terms of the
$\lambda$ and $\omega$ parameters. By using eqs. (\ref{size}),
and (\ref{jump3}), we obtain the result
\be
\frac{3 e^{\lambda \theta_{min}}}{2\kappa^2_6\,b_0}=|\Lambda_i|\,,
\ee
where $|\Lambda_i|$ stands for the absolute value of the brane tensions.
The above relation fixes the physical size of the longitudinal extra dimension,
which constitutes a fundamental parameter of the model, in terms of the
brane tensions. The logarithmic dependence on the value of $|\Lambda_i|$
ensures the smallness of the size of the longitudinal dimension even for
large values of the brane tension. It is therefore the
existence of this extra dimension that introduces an additional parameter
in the model whose fixing replaces the fine-tuning between bulk and
brane parameters. Note, however, that the locations of the two 4-branes along
the transverse $\varphi$-dimension and, thus, the size of this
dimension remain a free parameter. We address this point in the next
section.

A final comment on the number of fine-tunings is in order at this point:
had the solutions for $\lambda<0$ led to a consistent spontaneous
compactification of the $\theta$-dimension, an additional problem would
appear: the special values of the $\omega$ parameter, for which these
solutions arise, would result in the fixing of the values of both brane
tensions through (\ref{jump3}). The fine-tuning between brane tensions and bulk
cosmological constant would still be absent however the number of necessary
fine-tunings in the model would be again two. We might therefore conclude
that periodic solutions, which respect the assumptions of our model and,
at the same time, demand less fine-tuning than the usual 5D brane models,
cannot arise in the framework of our analysis.


\section{Non-static 4D solutions}

Next, we proceed to construct 6-dimensional solutions with a
4-dimensional time-dependent submanifold but with a constant
radion field once again. 
The source of this time-dependence will be a non-vanishing 4D
effective cosmological constant.  Solutions similar to these but in the
presence of only one extra dimension have been derived before \cite{bent}.
Here, we will investigate the possibility whether such solutions
arise in the case of one additional, extra, longitudinal dimension. As
in the previous section, we are going to assume that, initially, both
extra dimensions are non-compact with the size of the $\varphi$-dimension
becoming finite due to the presence of the two 4-branes while the 
$\theta$-dimension will be appropriately warped and thus spontaneously
compactified. 

Going back to the full time-dependent Einstein's equations
(\ref{geq00})-(\ref{geq56}), we try once again the factorized ansatz:
$a(t, \theta, \varphi)=T(t)\,\Theta(\theta)\,\Phi(\varphi)$. The
off-diagonal component (\ref{geq56}) gives, as in the static case,
a proportionality relation, $\Theta(\theta)=A_\theta\,f(\theta)$,
between the two $\theta$-dependent  functions of the metric tensor.
The remaining off-diagonal equations , eqs. (\ref{geq05})-(\ref{geq06}),
in conjunction with the above factorized ansatz, both lead to the 
result $b=const.$, that guarantees the staticity of the extra 
2-dimensional spacetime. 

Subtracting eq. (\ref{geq66}) from eq. (\ref{geq55}), we recover one
of the two equations that determine the solutions for the metric
functions $\Phi(\varphi)$ and $f(\theta)$, namely
\be
\frac{\Phi''}{\Phi}=f f''- f'^2=\omega^2\,.
\ee
Before trying to derive the second equation, we need to determine
the solution for the time-dependent function $T(t)$. Taking the sum
of eqs. (\ref{geq00}) and (\ref{geqii}), we find the result
\be
\frac{\dot T}{T}\,\Bigl(\frac{2 \dot T }{T} - \frac{\ddot T}{\dot T}
\Bigr)=0\,.
\label{time}
\ee
Obviously, one solution of the above equation is the trivial one,
$\dot T =0$, which leads to the static case of the previous section
with the Minkowski-like 4D submanifold. Clearly, the model accepts
another solution that may be determined by demanding the expression
inside brackets to be zero. Then, we obtain the alternative solution
\be
T(t)=\frac{1}{c_0\,(t-t_0)}\,,
\label{Tsol}
\ee
where $c_0$ and $t_0$ are integration constants. If we pass from the
conformal time $t$ to the physical time $\tilde t$, through the
transformation $d\tilde t=T(t)\,dt$, we might easily see that
the solution takes the form
\be
T(\tilde t)=e^{H\,(\tilde t-\tilde t_0)}\,.
\label{Tsol2}
\ee
In the above expression, $H = c_0$ is again an integration constant, which
may be either real or imaginary describing a de Sitter
\be
ds_4^2=-d\tilde t^2 + e^{2H\tilde t}(dx^2 + dy^2 + dz^2)
\ee
or Anti de Sitter
\be
ds_4^2= dx^2 + e^{2Hx}(-d\tilde t^2 + dy^2 + dz^2)
\ee
4D submanifold, respectively. Since the Einstein's equations are all expressed
in terms of the conformal time, we will continue using $t$, instead of
$\tilde t$, and the form of the solution (\ref{Tsol}), instead of (\ref{Tsol2}),
for convenience. In terms of the conformal time, a real or imaginary $c_0$
will distinguish between a de Sitter or Anti de Sitter 4D submanifold.

Having determined the solution for $T(t)$, we now turn to eq. (\ref{geq66}),
which can be brought to the form
\be
6\Bigl(\frac{\Phi'}{\Phi}\Bigr)^2 -6 \frac{b^2 c_0^2}{A_\theta^2 \Phi^2}
+ 6f'^2 + 4 f f'' = -\kappa^2_2\, b^2 f^2 \Lambda_B\,.
\ee
Comparing the above with eq. (\ref{rear2}), we may easily see that we can
recover the equation (\ref{eqf2}) for $f(\theta)$, if $\Phi(\varphi)$
satisfies the following equation
\be
\Bigl(\frac{\Phi'}{\Phi}\Bigr)^2 - \frac{b^2 c_0^2}{A_\theta^2 \Phi^2}
-\omega^2=0\,.
\label{timephi}
\ee
The solution for the $\varphi$-dependent part of the warp factor, now,
takes the form
\be
\Phi(\varphi)= 
\frac{b\,c_0}{A_\theta\,\omega}\,
\sinh[\omega\,|\varphi-\varphi_0|]\,,
\label{dS}
\ee
for de Sitter space-time, and
\be
\Phi(\varphi)= \frac{b\,{\rm Im}(c_0)}{A_\theta\,\omega}\,\cosh[\omega\,
(\varphi-\varphi_0)]\,,
\label{AdS}
\ee
for Anti de Sitter space-time.

Since the basic equation for the metric function $f(\theta)$ has remained
unchanged, the general solution given by eqs. (\ref{general}) and
(\ref{lambda}) still remains the same. The subsequent discussion
on the possible ways of compactifying the $\theta$-dimension,
in the case of negative or positive bulk cosmological constant,
and the main conclusions drawn at the end of section 3 still hold. 
On the other hand, the time-dependence of the warp factor has radically
changed the solution for its $\varphi$-dependent part. Instead of
the monotonic exponential dependence that prevailed in the static
case, the time-dependent case may accommodate both $sinh$ and
$cosh$-type solutions. Solutions similar to the above usually
arise in the case where classical \cite{kop1} or quantum \cite{hkp}
effects from bulk scalar fields are taken into account or when
the effect of a bulk stabilizing potential is included in the
model \cite{kop1,kop2}. Note, however, that no such effects have been
assumed to be present in our analysis. The conclusions drawn in the previous
section from the {\it jump} conditions are also likely to change.
Substituting the above solution for $\Phi(\varphi)$ into the
{\it jump} conditions (\ref{jump1}), we obtain the constraints
\be
\omega\,\coth[\omega\,(\varphi_i-\varphi_0)] = (-1)^i 
\,\frac{\kappa^2_6}{3}\,b_0\,\Lambda_i
\label{jumpt1}
\ee
for the $sinh$-type solutions, and
\be
\omega\,\tanh[\omega\,(\varphi_i-\varphi_0)] = (-1)^i
\,\frac{\kappa^2_6}{3}\,b_0\,\Lambda_i
\label{jumpt2}
\ee
for the $cosh$-type solutions. In the above two equations,
$i=1,2$ denotes again the two 4-branes. The $sinh$-type solution with
its monotonic behaviour can clearly accommodate only pairs of
positive-negative tension branes in analogy with the static
case. Moreover, the specific solution is plagued by the existence
of a singularity at the point $\varphi=\varphi_0$. In order to
exclude the singularity from the 6D spacetime, both branes
must be located on the same size of the singularity, i.e.
$\varphi_1,\varphi_2<\varphi_0$. On the other hand, the
$cosh$-type solution is everywhere well defined. The same point,
$\varphi=\varphi_0$, corresponds, in this case, to a minimum
which allows for pairs of positive tension branes to fit in the
model. In both cases, either for $sinh$ or $cosh$-type solutions,
the above {\it jump} conditions will fix the location of the two
branes relative to the singularity or the minimum,
respectively, in terms of the two brane tensions. Note, however,
that the brane tensions, considered as input parameters of
the model, remain totally uncorrelated. Moreover, the lack of
fine-tuning between any of the brane tensions and the bulk cosmological
constant still holds rendering the model free of any fine-tuning.


\section{Radion dynamics}

The main goal of the previous sections was to find 6D solutions, static
and non-static, with a constant radion field which demand less
or no-fine-tuning of their parameters compared to other models in the
literature. Nevertheless, an important 
aspect of these solutions needs to be studied next: do these solutions
actually extremize the radion effective potential? And, if yes, is
this extremum a minimum or a maximum of the effective potential?

In Ref. \cite{GKL}, a 5-dimensional ``extremization'' constraint that
could serve as a consistency check for any solutions with a constant
radion field, was derived. It might be worth deriving the corresponding
constraint in 6 dimensions and comment on the possible differences
that arise as one changes the number of extra dimensions. In order
to do that, we need to go back to the time-dependent Einstein's equations,
and start by constructing the 4D trace of the energy-momentum tensor
(by taking the sum of eq. (\ref{geq00}) with three times eq. (\ref{geqii})),
which comes out to have the form
\ba
&~& \hspace*{-2cm} \kappa^2_6\,T^\mu_\mu = -\frac{4\dot b^2}{a^2 b^2}-
\frac{6 \ddot b}{a^2 b} -\frac{12 \dot a \dot b}{a^3 b} 
-\frac{6 \ddot a}{a^3}+ \frac{4 f''}{b^2 f} + \frac{12}{b^2 f^2 a^2}\,
\biggl(\frac{\partial a}{\partial \varphi}\biggr)^2 
\nonumber \\[4mm]
&~& \hspace*{2cm} + \,\frac{12}{b^2 f^2 a}\,\frac{\partial^2 a}
{\partial \varphi^2} + \frac{12 f'}{b^2 f a}\,\frac{\partial a}
{\partial \theta} + \frac{12}{b^2 a^2}\,\biggl(\frac{\partial a}
{\partial \theta}\biggr)^2 + \frac{12}{b^2 a}\,\frac{\partial^2 a}
{\partial \theta^2}\,.
\label{con1}
\ea
Next, we construct the following linear combination of all the components
of the 6D energy-momentum tensor
\ba
&~& 
\hspace*{-2cm} 
\kappa^2_6\,\Bigl[(4-n)\,T^\mu_\mu + 3(n-2)\,T^5_5
+ 3(n-2)\,T^6_6\Bigr]= 
\nonumber \\[4mm] 
&~& -\biggl[(4-n)\,\frac{4\dot b^2}{a^2 b^2}+
\frac{12 \ddot b}{a^2 b} +\frac{24 \dot a \dot b}{a^3 b}\biggr]
-12 (n-1)\,\frac{\ddot a}{a^3} + (4-n)\frac{4 f''}{b^2 f}
\nonumber \\[4mm] 
&~& +\frac{24}{b^2}\,\biggl[ \frac{(n-1)}{f^2 a^2}\,
\biggl(\frac{\partial a}{\partial \varphi}\biggr)^2
+ \frac{1}{f^2 a}\,\frac{\partial^2 a}{\partial \varphi^2} 
+ \frac{f'}{f a}\,\frac{\partial a}{\partial \theta} + 
\frac{(n-1)}{a^2}\,\biggl(\frac{\partial a}{\partial \theta}\biggr)^2
+ \frac{1}{a}\,\frac{\partial^2 a}{\partial \theta^2}\biggr]\,,
\label{cona}
\ea
where $n$ is an integer. In the case where the warp factor can be written
as the product of two functions, one depending on the 4D coordinates and one
on the extra coordinates, the above expression can be greatly simplified.
If we write the warp factor in the factorized form
\be
a(t,\theta,\varphi)=T(t)\,W(\theta,\varphi)\,,
\label{fact}
\ee
the expression in the last line of eq. (\ref{cona}) may be written as
\ba
&~& \hspace*{-2cm} \frac{24}{b^2}\,\biggl[ \frac{(n-1)}{f^2 a^2}\,
\biggl(\frac{\partial a}{\partial \varphi}\biggr)^2
+ \frac{1}{f^2 a}\,\frac{\partial^2 a}{\partial \varphi^2} 
+ \frac{f'}{f a}\,\frac{\partial a}{\partial \theta} + 
\frac{(n-1)}{a^2}\,\biggl(\frac{\partial a}{\partial \theta}\biggr)^2
+ \frac{1}{a}\,\frac{\partial^2 a}{\partial \theta^2}\biggr]
\nonumber \\[3mm]
&~& \hspace*{2cm} 
=\frac{24}{n W^n}\,\frac{1}{\sqrt{g^{ex}}}\,\partial_m
\Bigl[\sqrt{g^{ex}}\,\partial^m W^n(x^m)\Bigr]=
\frac{24}{n W^n}\,(D_m D^m W^n)\,,
\ea
where $x^m$ denotes the extra coordinates ($\theta, \varphi$) and
$g^{ex}_{mn}$ the metric tensor of the 2-dimensional extra spacetime.
The above 2D double covariant derivative of the function $W^n$ divided by
$W^n$ is the analogue of the double derivative of the same function
appearing in the 5D constraint of Ref. \cite{GKL}. In the more general
case where more than one, possibly non-flat, extra dimensions are present
in the theory, the
double derivative needs to be replaced with the covariant derivative
in order for the internal geometry of the extra spacetime to be taken
into account. 

In addition, we may rewrite some of the other terms appearing in eq. (\ref{cona})
in the following way
\be
12\frac{\ddot a}{a^3}= \frac{2}{W^2}\,R_{(4)}\,, \qquad \qquad
\frac{4 f''}{b^2 f}=-2 R_{(2)}\,,
\ee
where $R_{(4)}$ and $R_{(2)}$ stand for the 4D and 2D, respectively,
scalar curvature. Then, the constraint (\ref{cona}) takes the simplified
form
\ba
&~& \hspace*{-1cm} \frac{24}{n W^n}\,(D_m D^m W^n)=
\kappa^2_6\,\Bigl[(4-n)\,T^\mu_\mu + 3(n-2)\,T^5_5 + 3(n-2)\,T^6_6\Bigr]
\nonumber \\[3mm]
&~& \hspace*{1cm} +\,(n-1)\,\frac{2}{W^2}\,R_{(4)} + 2 (4-n)R_{(2)} +
\biggl[(4-n)\,\frac{4\dot b^2}{a^2 b^2}+
\frac{12 \ddot b}{a^2 b} +\frac{24 \dot a \dot b}{a^3 b}\biggr]\,.
\label{conb}
\ea
Compared to the 5-dimensional one \cite{GKL}, the 6-dimensional version of
the above constraint has a similar but more generalized structure. It
involves all the extra components of the energy-momentum tensor,
namely $T^5_5$ and $T^6_6$, as anticipated, and both on an equal footing.
Moreover, the 2-dimensional scalar curvature of the extra spacetime explicitly
makes its appearance together with the 4-dimensional one. Finally, the
coefficients appearing in front of the 4D trace and extra components of
the energy-momentum tensor seem to be ``dimension''-dependent. By mere
comparison of the 5D and 6D versions of the constraint, we may easily
conclude that the coefficient in front of the extra components 
behaves as $(1+d)(n-2)$, where $d$ the number of extra dimensions. 
However, the dependence of the coefficient of the 4D trace is more subtle
and a higher dimensional calculation could only reveal its exact form
in terms of $d$.

We would also like to stress here an additional point on the form of the
constraint (\ref{conb}). It holds for every solution of the 6D Einstein's
equations, even for the ones with a non-static extra spacetime. For the
particular case of $n=1$ the above constraint can be interpreted as the
equation of motion of the time-dependent scalar field $b(t)$. This can
become clear if we rewrite Eq. (\ref{conb}) as
\be
\frac{2}{b^2}\,D_\mu D^\mu b^2= \kappa^2_6\,(T^\mu_\mu - T^5_5 -T^6_6)
+ 2 R_{(2)} -\frac{8}{W}\,(D_m D^m W)\,.
\label{eqb}
\ee

For solutions with a non-static extra spacetime, the {\it rhs} of the 
above equation vanishes when an extremum, either minimum or maximum,
is reached. For solutions with a constant radion field, which by
definition correspond to an extremum of the radion effective potential,
the same combination should also vanish. Therefore, the solutions
found in the two previous sections, either static or non-static, but
with constant $b$, should satisfy the above constraint. By using the
following expressions for the components of the energy-momentum tensor
\be
T^\mu_\mu= -4\Lambda_B -\frac{4 \Lambda_i}{b_0 f^2}\,\delta(\varphi-\varphi_i)\,,
\qquad 
T^5_5= -\Lambda_B -\frac{\tilde \Lambda_i}{b_0 f^2}\,\delta(\varphi-\varphi_i)\,,
\qquad T^6_6=-\Lambda_B\,,
\ee
setting $W \equiv \Theta(\theta)\,\Phi(\varphi)$ and employing the form
of the solutions found in sections 3 and 4, we obtain
\be
- 2\,\biggl(\kappa^2_6\Lambda_B
+10\,\frac{\lambda^2}{b_0^2}\biggr) -\frac{8\delta(\varphi-\varphi_i)}{b_0 f^2}\,
\biggl[\frac{[\partial_\varphi a]_i}{b_0\,a} +
\frac{\kappa^2_6}{3}\,\Lambda_i\biggr]\,.
\ee
By using the {\it jump} conditions (\ref{jump1}) and the definition of
the $\lambda$ parameter from eq. (\ref{lambda}), we may easily conclude
that both the static and non-static solutions derived in the previous sections
satisfy the ``extremization'' constraint, as anticipated.

In order to answer the question whether the above extremum is a minimum or
a maximum, a perturbation analysis around the above solutions needs to
be performed, in which the time-dependent, small perturbation will be
associated with the radion field. Therefore, we consider the following
ansatz for the line-element of the 6-dimensional spacetime
\ba
&~& \hspace*{-2cm} ds^2=[a_0^2 f(\theta)^2 \Phi(\varphi)^2 + \epsilon\, 
A(\theta, \varphi)\,b(\tilde t)]\,(-d\tilde t^2 + e^{2H\tilde t}
d\vec{x}^2)\nonumber \\[2mm]
&~& \hspace*{3cm} +\, [1+\epsilon\,B(\theta, \varphi)\,b(\tilde t)]\,
(d\theta^2 + f(\theta)^2 d\varphi^2]\,.
\label{pert}
\ea
Note that we have switched to the system of non-conformal coordinates
and we have set $b_0=1$, for simplicity. We have also chosen to
perturb the de Sitter solutions found in the previous section, however,
our analysis can be easily extended to the cases of Minkowski
or Anti de Sitter branes by setting $H^2 \rightarrow 0$ or
$H^2 \rightarrow -H^2$ (with the $\tilde t$-dependent perturbations
replaced by $x$-dependent), respectively.

The above ansatz when substituted into the 6D Einstein's equations
will lead to a system of differential equations and constraints that
govern the behavior of the new line-element. We will work in the 
linear order approximation and, thus,  we are keeping only terms proportional
to the small parameter $\epsilon \ll 1$. In this approximation,
the generalized (00), ($ii$), (05) and (06)-components of
Einstein's equations take the form
\ba
&~& \hspace*{-1.7cm} \biggl(B + \frac{A}{a_0^2 f^2 \Phi^2}\biggr)\,3H \dot b -
(\kappa^2_6 a_0^2 f^2 \Phi^2\Lambda_B -3H^2)\,B b -
\frac{Ab f''}{f} - \frac{3b}{2f^2}\,\frac{\partial^2 A}{\partial
\varphi^2} \nonumber \\[3mm] &~&
\hspace*{0.7cm} - \frac{3b f'}{2f}\,\frac{\partial A}{\partial \theta}-
\frac{3b}{2}\frac{\partial^2 A}{\partial \theta^2} -
\frac{a_0^2b \Phi^2}{2}\,\biggl(\frac{\partial^2 B}{\partial 
\varphi^2} + ff'\,\frac{\partial B}{\partial \theta}
+f^2 \frac{\partial^2 B}{\partial \theta^2}\biggr) =
\kappa^2_6 A b\,\Lambda_B\,,
\label{pert1} \\[5mm]
&~&\hspace*{-1.7cm} \biggl(B + \frac{A}{a_0^2 f^2 \Phi^2}\biggr)\,(-2H \dot b
- \ddot b) + (\kappa^2_6 a_0^2 f^2 \Phi^2 \Lambda_B -3H^2)\,B b +
\frac{Ab f''}{f} + \frac{3b}{2f^2}\,\frac{\partial^2 A}{\partial
\varphi^2}\nonumber \\[3mm] &~&
\hspace*{0.4cm} + \frac{3b f'}{2f}\,\frac{\partial A}{\partial \theta} +
\frac{3b}{2}\frac{\partial^2 A}{\partial \theta^2} +
\frac{a_0^2b \Phi^2}{2}\,\biggl(\frac{\partial^2 B}{\partial
\varphi^2} + ff'\,\frac{\partial B}{\partial \theta}
+f^2 \frac{\partial^2 B}{\partial \theta^2}\biggr) =
-\kappa^2_6 A b\,\Lambda_B\,,
\label{pert2} \\[5mm]
&~& \hspace*{2.9cm} \biggl(2B+\frac{3A}{a_0^2 f^2 \Phi^2}\biggr)\,\frac{f'}{f}
-\frac{3}{2 a_0^2 f^2 \Phi^2}\,\frac{\partial A}{\partial \theta}
-\frac{1}{2}\,\frac{\partial B}{\partial \theta} = 0\,,
\label{pert3} \\[5mm]
&~& \hspace*{2.8cm} \biggl(2B + \frac{3A}{a_0^2 f^2 \Phi^2}\biggr)\,
\frac{\Phi'}{\Phi}
-\frac{3}{2 a_0^2 f^2 \Phi^2}\,\frac{\partial A}{\partial \varphi}
-\frac{1}{2}\,\frac{\partial B}{\partial \varphi} =0\,,
\label{pert4}
\ea
respectively. 

By taking the sum of the (00) and ($ii$)-components, we end up with the
constraint
\be
\biggl(B + \frac{A}{a_0^2 f^2 \Phi^2}\biggr)\,
(H \dot b - \ddot b)=0\,,
\label{options}
\ee
which demands that one of the two expressions inside brackets vanishes.
If we assume that the expression inside the first bracket is zero, then, we
can determine the exact form of the unknown functions $A(\theta, \varphi)$
and $B(\theta, \varphi)$ by plugging this constraint into the 
off-diagonal components (\ref{pert3})-(\ref{pert4}). Then, we
find that
\be
A(\theta, \varphi)=1\,, \qquad B(\theta, \varphi)=-\frac{1}
{a_0^2 f(\theta)^2 \Phi(\varphi)^2}\,,
\label{solAB}
\ee
modulo a multiplicative constant. Both of the equations
(\ref{pert1})-(\ref{pert2}), in that case, reduce to the
background equation (\ref{timephi}) (with $c_0^2$ being replaced
by $H^2$ due to the use of non-conformal coordinates) which is
obviously satisfied by the background solution.
However, the solution (\ref{solAB}) fails to satisfy any of the
remaining components of Einstein's equations. For example, the
off-diagonal (56)-component, which has the form
\be
\frac{2f'}{f}\,\biggl(\frac{\partial A}{\partial \varphi}-
A\,\frac{\Phi'}{\Phi}\biggr) + a_0^2 f^2 \Phi^2\biggl(
\frac{f'}{f}\,\frac{\partial B}{\partial \varphi} +
\frac{\Phi'}{\Phi}\,\frac{\partial B}{\partial \theta}\biggr)
+\frac{\Phi'}{\Phi}\,\frac{\partial A}{\partial \theta}-
\frac{\partial^2 A}{\partial \theta \partial \varphi}=0\,,
\label{pert7}
\ee
leads to the constraint $f' \Phi'=0$ which is obviously in contradiction
with the form of the background solution. Finally the (55) and 
(66)-components, which after using eq. (\ref{solAB}) take the simplified form
\ba
&~& \frac{(D_\mu D^\mu b)}{a_0^2 f^2 \Phi^2} -
\frac{12b}{f^2}\,\bigg(\frac{\Phi'^2}{\Phi^2} -
\frac{H^2}{a_0^2 \Phi^2}\biggl) -\frac{12 b f'^2}{f^2}
-\frac{4b \Phi''}{f^2 \Phi}=\kappa^2_6\,b\,\Lambda_B\,, \\[4mm]
&~& \frac{(D_\mu D^\mu b)}{a_0^2 f^2 \Phi^2} -
\frac{b}{f^2}\,\bigg(\frac{8\Phi'^2}{\Phi^2} -
\frac{12 H^2}{a_0^2 \Phi^2}\biggl) -\frac{12 b f'^2}{f^2}
-\frac{4b f''}{f}=\kappa^2_6\,b\,\Lambda_B\,,
\ea
respectively, also turn out to lead to an inconsistent result. As a result,
the option that the first expression inside brackets in eq. (\ref{options})
is zero needs to be excluded. We are therefore left with the alternative
option $(H \dot b- \ddot b)=0$ which leads to an exponential solution
of the time-dependent perturbation. By employing the equation of motion
of the ``radion'' field
\be
g^{\mu\nu} D_\mu D_\nu b = m^2 b
\ee
and the exponential form of its solution, we get the result 
$m^2=-4H^2$. Obviously, the mass squared of the radion field
turns out to be negative for de Sitter 4D subspace, zero for
Minkowski and positive for Anti de Sitter 4D subspace. 

The above result is in perfect agreement with similar works conducted in
5 dimensions: in Refs. \cite{Gen,GKL, AP} (see also \cite{BDB}), it has
been shown
that the system of two Minkowski branes with a zero total brane tension
cannot be stabilized since it leads to a radion field with a zero mass.
In that case, the extremization of the radion potential follows from
the fine-tuning of the parameters of the model which however cannot
create a unique minimum for the radion field in the absence of a physical
stabilization mechanism. As a result, the static solution derived in
Section 2 describing a pair of flat branes cannot be stabilized and
the radion field remains always massless.
The same conclusion was drawn in Refs. \cite{BDL, CF} where the
stability of curved branes was also studied. According to their results,
a pair of de Sitter branes leads to an effective theory for the radion
field with a negative mass squared while a pair of Anti de Sitter
branes turns out to be stable since it leads to a positive radion mass
squared. Therefore, our solution (\ref{dS}) corresponding to two
branes with a positive effective cosmological constant is unstable
under small time-dependent perturbations while the
alternative solution described by (\ref{AdS}) and corresponding to
two branes with negative effective cosmological constant is stable.
We bear in mind that this conclusion ignores any additional contribution to
the 4D radion potential as might be expected from a more complete model
which includes supersymmetry and supersymmetry breaking. Let us 
finally note that the above results are also in agreement with those
of Ref. \cite{zhuk} where the extremum of the radion effective potential
was a global minimum only in the case of a 4D negative cosmological
constant.


\section{Conclusions}

In this paper, we have presented a 6D brane world model in the framework of which
we addressed a number of important issues that arise in the context of higher-dimensional
brane models. The relaxation of the fine-tuning of the fundamental parameters of the
model was one of them: solutions with less or no fine-tuning at all, compared to
their 5D analogues, were constructed where the severe correlation between brane
and bulk parameters was replaced by the fixing of the sizes of the two extra dimensions.
In this way, another problem, that of the determination of the inter-brane distance
was automatically resolved. The solutions derived allow for the introduction of
pairs with positive-negative brane tensions, in analogy with the RS model, however
more physically interesting configurations with pairs of only positive tension branes
were also found. Finally, the issue of the behaviour of the aforementioned solutions,
under time-dependent perturbations around configurations with a constant radion
field, was examined and conclusions regarding their stability were derived. 

In more detail, 
considering a 2D internal space of constant curvature, an exact static solution 
was first presented. The warp factor exhibits neither spherical nor cylindrical
symmetry but depends on both extra coordinates. Two 4-branes are introduced at
two different points along the so-called ``transverse'' extra dimension, along
which the warp factor is a pure exponential resembling the profile of the warp
factor in the case of the 5D RS model. The remaining extra dimension, the 
``longitudinal'' one, along which the 4-branes extend, was shown to be 
``spontaneously'' compactified in the case of a negative bulk cosmological
constant, or equivalently of a negatively curved internal space. Although
the consistency of the bulk solution with the brane boundary conditions demand
the two branes to have exactly equal and opposite tensions, no correlation
exists between the brane tensions and the bulk cosmological constant.  
Instead, it is the size of the longitudinal extra dimension that is fixed
through the {\it jump} conditions in terms of the value of the brane tensions.

The above solution, although it is characterized by reduced fine-tuning of its
fundamental parameters, has the inter-brane distance along the transverse
extra dimension as a free parameter. In an attempt to resolve this problem, too,
we then derived a non-static solution whose 4D subspace corresponds to a 
de Sitter or Anti de Sitter spacetime. These solutions have a number of positive
features: first, being non-static due to the presence of an effective 4D cosmological
constant, the only fine-tuning of the model, the one between the two brane tensions,
is also relaxed rendering the model free of any fine-tuning; second, although they
have the same profile along the compactified,
longitudinal dimension, the exponential behaviour along the transverse one changes
to a $cosh$ or $sinh$-type one, a fact which has two important consequences: the 
fixing of the locations of the two branes and, thus, of the inter-brane distance,
and the accommodation of pairs of positive tensions branes instead of only
pairs of positive-negative tensions.  

The fixing of the physical inter-brane distance and thus of the size of the
extra dimension relies on the assumption that the scale factor along the internal
spacetime remains constant. Both of the above solutions, static or non-static,
were derived under this assumption, or, alternatively, that the corresponding 
``radion'' field was already at the extremum of its effective potential. An
``extremization'' constraint, valid for solutions with static or non-static
radion field whose effective potential possess an extremum, was formulated
and used as a consistency check for our solutions. The
final issue of whether this extremum was a unique minimum or merely a local
maximum of the radion effective potential needed to be addressed. The
background solutions with a constant scale factor along the extra dimensions
were perturbed and a linear stability analysis was performed. Our analysis
revealed the stability of the non-static solutions describing an Anti de
Sitter 4D spacetime and accommodating pairs of only positive tension branes.
The remaining solutions describing 4D de Sitter and Minkowski spacetimes
were found to correspond to local maxima and saddle points, respectively,
in close analogy to similar analyses performed in 5 dimensions. 

\bigskip

\noindent{\Large \bf Acknowledgements}\,

\noindent
We are deeply grateful to Luigi Pilo, Maxim Pospelov and Alexander
Zhuk for useful discussions. 
P.K. would also like to acknowledge financial support by the DOE Grant No. 
DE-FG-02-94-ER-40823 and by the EC TMR contract No. HRPN-CT-2000-00148
during the early and late stages of this work, respectively, as well
as the CERN Theory Division for the hospitality and financial support
while this work was in progress. 
R.M. wishes to thank the CERN Theory Division for its hospitality during
a portion of this work.
The work of K.O. was supported in part by  DOE Grant No. 
DE-FG-02-94-ER-40823 at the University of Minnesota.

\end{document}